# Defense Strategies Against Modern Botnets

Srdjan Stanković
Personnel Department
Ministry of Defense
Belgrade, Serbia
srdjan.stankovic@mod.gov.rs

Dejan Simić
Department of Information Technology
Faculty of Organizational Sciences
Belgrade, Serbia
dejan.simic@fon.rs

*Abstract*—Botnets are networks of compromised computers with malicious code which are remotely controlled and which are used for starting distributed denial of service (DDoS) attacks, sending enormous number of e-mails (SPAM) and other sorts of attacks. Defense against modern Botnets is a real challenge. This paper offers several strategies for defense against Botnets with a list and description of measures and activities which should be carried out in order to establish successful defense. The paper also offers parallel preview of the strategies with their advantages and disadvantages considered in accordance with various criteria.

## I. INTRODUCTION

IT security involves a lot of fields nowadays and they are focused on various security aspects, starting form the lowest layers of OSI models up to applicative ones. Since security at lower layers has been significantly improved, the attackers have redirected their interest towards the highest layers of OSI models. In most cases they try to find the entrance to the systems via applicative layer. Having in mind the fact that fast-spreading Internet has been stimulated by the services which provide conditions for exchanging and updating of the information independently from software platforms, question of security on the Internet has become much more important.

Large number of the Internet attacks happening nowadays is directed toward exploitation of individuals and organizations in order to earn money, and that often causes financial losses. One of the most serious threats to the Internet is presence of large number of compromised computers. Networks of such computers are present everywhere. They are mostly controlled by one or a few hackers and are used for different types of attack – starting from Distributed Denial-of-Service (DDoS), sending of unwanted e-mail messages (SPAM) up to spreading of malicious programs etc [1], [3], [4]. Unlike other types of attacks, attacks performed by Botnets which usually consist of several hundred of computers, can collect large amount of computer resources and exploit them for performing of various kinds of attacks. That's why hackers are especially interested in their usage in order to reach maximum of benefits. In the same time, harm caused by usage of such networks is matchlessly bigger than the one caused by traditional, discrete attacks. Whole Internet community, legislative institutions, individual users and big companies have been considering possibilities to confront this problem which is one the most serious security threats directed against the Internet community today. The available literature contains only a few information about defense against Botnets [5], [16], [17], [21] and the information deal with only specific aspects of defense.

Since the available literature does not consider total defense against Botnets sufficiently, the second section of this paper describes Botnets and the ways of their functioning. The third section deals with the problems caused by usage of Botnets, while the fourth section considers and describes types of Botnets. The fifth section offers preview of the most common attacks produced by Botnets as well as their description. In the sixth, central section, one can find strategies of defense against Botnets, i.e. list of the measures and activities that should be carried out in order to protect successfully the system against Botnets. Parallel preview of the strategies with their advantages and disadvantages considered according to various criterions can also be found in this section.

## II. PROBLEM STATEMENT

An infected computer-zombie, while carrying out malicious code, spends the resources and obeys certain commands without permission and knowledge of the owner and that activity causes slowing down of the computer, showing of some mysterious messages or it can even cause collapse of the system. [3]. However, this is not the biggest problem.

The problem with Botnets appears when they are used for attack. Botnet of a million robots, with uploading speed of 128 Kbps per infected computer (zombie), can reach size of 128 gigabits in traffic. It is enough to put out of function 500 companies and several countries by applying DDoS attacks. If several big Botnets unite, they could threaten functioning of national infrastructure of the most countries [5].

From statistical point of view, Botnets which generate SPAM participate a lot in distribution of SPAM. Total level of SPAM has reached 82.7% in February 2008, and it reaches average level of 81.2% in a year, compared to 84.6% in 2007. Approximately 90% of SPAM is distributed by means of Botnets, including disreputable Botnets called storm. [6].

## III. BOTNETS

The expression Botnet was created by joining the words BOT –short form of Robot and NET. Botnet is a group of computers infected with a kind of malicious robot software and the computers are very dangerous for security of the owner's computer. After the robot software has successfully been installed into the owner's computer, the computer becomes a zombie [7]. It carries out the orders given by Bot controller (Botmaster) without agreement and approval given by the owner.

Botnet can be either small or big, depending on complexity and sophistication of the robot which is used. A big Botnet can be composed of 10,000 individual zombies. On the other hand, a small Botnet can be composed of only a thousand drones. The owners of zombie computers usually do not know that their computers are remotely controlled and exploited by some individual or a group of disreputable programs, and the remote control is usually carried out by Internet Relay Chat (IRC) protocol.

Technology for remote monitoring of compromised computers appeared for the first time at the end of 1999, when the researchers from Sans Institute detected the remote executive code in thousands of Windows computers. The infected computers were called robots because of the very nature of remote control and soon after that the expression was shortened to Bot. The researchers could not predict what the code would do because the code was crypto protected, but four months later, in February 2000, the robot caused DDoS attacks and the attacks alternatively made inaccessible Amazon, eBay and other sites for electronic purchase in a week period [1].

Zombie computer is created after infection by virus or a worm which opens TCP communication port through which Trojan program is unauthorizedly inserted into the computer. This program is activated depending on the need in order to perform some of the tasks such as sending of SPAM mail, further distribution of other viruses or DDoS attacks against some Web site or DNS server.

Botnet network is formed for commercial reasons, either in order to earn money by sending SPAM mail or tampering of rivals. The biggest problem which the organizer can have is Command and Control (C&C) of the net [8]. The same as many other program tools based on network communication, Bot programs also mutually communicate with their owners by means of well defined network protocols. In most cases Botnet nets do not create new net protocols, but they use already existing, implemented by the tools which are available in public (for example IRC protocol as well as publicly available implementations of IRC clients and servers).

Besides Botnets which mostly use centralized commanding and control, there are Botnets which use Peer-to-Peer control, and in that case there is not a central controller (Botmaster) [2]. After the initial message that the attack begins is sent, these Botnets create "chain reaction" causing "Storm" and that's why these Bonnets are called "Storm" [9].

*A. IRC protocol*

Since it is easier to implement Botnets based on IRC protocol, it is not strange that this protocol is the most popular and mainly used for Botnet communication. This protocol is mostly developed for communication in groups (many-to-many), for discussion forums called "channels", but it also supports one-to-one communication by private messages. This is very useful for hackers because it is possible for them to send the orders to whole Bot net of computers or only to some of them by sending private messages.

Botnet C&C server has got initiated IRC service which does not differ at all from the standard IRC ones [10]. In most cases Botmaster creates certain IRC channel on server, and all Bot computers are connected to it waiting for the orders (Figure 1.). Inspection of IRC net traffic can detect presence of Botnets in local network, since usage of IRC client/server is usually not allowed in corporative networks.

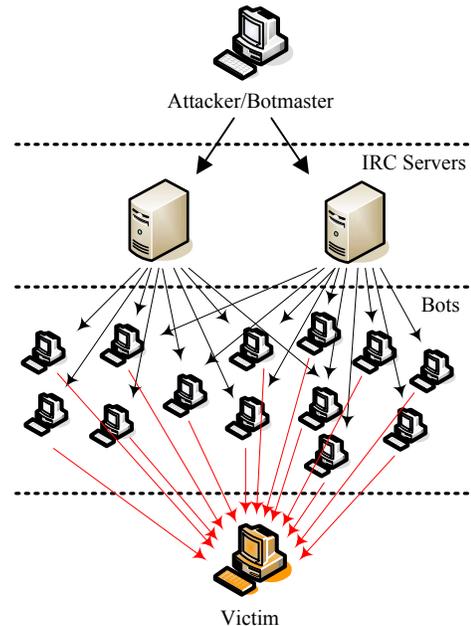

Figure 1. Elements of Typical IRC BotNet

This misusage of IRC has caused the situation in which most of conventional IRC systems block the access from already known Botnets, so that "controllers" must look for some new or even form their own illegal IRC servers.

*B. HTTP protocol*

It is not obligatory for a zombie computer to use IRC control system. It can use Web access via compromised Web server. The point is that Trojan/Bot component starts the connection towards some ordinary Web server.

HTTP protocol is also one of popular methods of communication which is used in Botnet networks [11]. There are two basic advantages of usage of this protocol. On one hand, the network which is based on HTTP protocol can be hidden in a better way in the rest of Internet traffic while, on the other hand, administrators can rarely forbid complete HTTP traffic. This means that the network has a permanently open communication channel and it is in the same time one of the optimum techniques to bypass security rules.

Detecting of Botnet networks which use HTTP protocol for communication is much more difficult, because this sort of traffic usually interferes with high capacity of normal HTTP network traffic. That's why HTTP requirements are not noticed and they are approved by Firewall. However, development of appropriate filters makes their detection possible, since there is difference between that kind of traffic and normal HTTP traffic.

*C. Other protocols*

Some more developed Botnet networks use other protocols (IM protocols, P2P protocols) for communication. New versions of Botnet use P2P communication, especially crypted implementation of P2P protocol, which is designed for communication by means of private messages and exchanging of files among small number of confidential parties. Nevertheless, relatively small number of Botnet networks does not use IRC or HTTP. It is possible that other protocols will be used more in the future, and that would be a big challenge in development of some countermeasures.

## IV. TYPES OF BOTNET

There are more types of Botnet, however, modern Botnets are [12]:

- Agobot, Phatbot, Forbot, XtremBot; It is probably the best known Bot. More than 500 hundred versions of Bot have been registered so far, thanking to the fact that it has been written in C++ language and that the original code is accessible. It has Sniffer and Rootkit protection. Bot is structured in a good modular way and it is very easy to add some new commands to it. It is quite enough to expand CcommandHandler or Cscanner classes and add your own function. Besides, Agobot is the only Bot which uses not only IRC protocol but other protocols, too.

- SDBot/Rbot/UrBot; At this moment it is the most active group of IRC Bots. It is written in C language but in a bad way if we compare it to Agobot. Its characteristics are similar to the one of Agobot, although the group of the orders is not so big. It is very popular with the attackers.

- mIRC-based Bots – GT-Bots; There are many versions of these Bots on the Internet, mostly because mIRC is one of most often used IRC clients for Windows. GT stands for "global threat" and it is common name for the Bots written for usage of mIRC. GT Bots use mIRC client for starting of a group of binary files (mostly DLL files) and scripts.

Besides already mentioned three types of Bots which we can find every day, there are some other Bots which can be met very rarely. Some of the robots have features which make them worth of mentioning [12]:

- DSNX Bot; Dataspy Network X Bot is written in C++ language and it has got system of Plugins which enables further spreading, usage of DDoS attacks or hidden Web server.

- Q8 Bot; Q8 Bot is very small and it is written in less than 1.000 lines of C code exclusively for Unix/Linux platforms. It contains dynamic updating via HTTP protocol, various DDoS attacks, carrying out of the order arbitrarily, etc.

## V. MALICIOUS USES OF BOTNETS

Botnet is nothing but means or tool for performing of different types of attacks. You can find some kinds of the attacks further in the paper.

*A. Denial of Service Attacks*

Distributed Denial-of-Service (DDoS) attacks are often used by Botnet. DDoS attack is a kind of attack against computer systems or other networks which causes loss of the service for the user. Besides, it is possible to overload the communication channel if DDoS attack sends too many packages per second. TCP and UDP "flood" are mostly used [12]. DDoS attacks are not limited only to Web servers, but almost all kinds of service available on the Internet can be targets of the attacks [13].

*B. Spamming*

Some Bots often have capability to open Socks v4/v5 proxy, generic proxy protocol for TCP/IP network applications on the compromised computer. After SOCKS proxy is enabled, the computer can be used for impious tasks such as sending of SPAMs. Using thousands of Bots, the attacker can send large amount of electronic mail (SPAM). Some Bots have special function of collecting electronic addresses. Besides, they are used for sending phishing-mail [14].

*C. Traffic Monitoring (Sniffing traffic)*

Bot can function as a sniffing traffic in order to detect and intercept sensitive data which pass through infected machines [4]. The data these Bots look for are user names and passwords, although they can take over other security sensitive information. If there are more than one Bot in the computer, traffic monitoring can enable collecting information about other Bots.

*D. Keylogging*

Some Bots are adjusted in such a way that they install keylogger programs on the infected computers. With keylogger program, Bot controller can use the program for filtering which collects only the sequence typed before or after interesting keywords such as PayPal or Yahoo mail.

*E. Mass Identity Theft*

Combination of various functionalities of Bonnets can often be used for comprehensive stealing of identity. In this case, false message sent by e-mail (phishing mails) which resembles the legitimate one, demands from the user-possible victim, to go to certain site and confirm their private data. The false e-mail message has been sent by Bot in form of SPAM. The same Bots can host several false Web pages pretending as if they are the original ones and they can collect personal information relating the victims [12].

*F. Botnet Spread*

Bots are mostly used for spreading of new Bots. All Bots have got implemented mechanisms for taking over and execution of files by means of HTTP, FTP or by sending e-mail with Bot virus.

*G. Pay-Per-Click Systems Abuse*

Botnet can be used for financial profit by automatic clicking on Pay-per-Click Systems. The infected computer can be used for automatic clicking on site while browser is activated. The purpose of usage of Botnets in this case is earning of money, because amount of the earned money depends on number of clicks in certain period of time [4]. This way of using Botnet is relatively unusual, but from the attacker's point of view, the idea is not bad.

## VI. DEFENSES OF THE BOTNETS

Defense against Bots and Botnets is carried out by application of certain strategies, i.e. certain measures and activities. All the Internet users are responsible for defense, starting form home or business computer users, system administrators, developers, up to Web service/application administrators. The defense must be considered as a permanent and comprehensive process in which all the activities must be proactive. This is the only way to achieve good results and to protect computers, i.e. Web services/applications against the activities with bad intentions.

In the following table (table I) there is a preview of possible defense strategies against Botnets as well their short description.

TABLE I.  STRATEGIES OF DEFENSE AGAINST BOTNETS

| Strategy of defense | Description |
|---|---|
| Defense against infection by Bot (DAIBB) | Don't let Botmaster insert Bot into computer |
| Defense against attack by Bot (DAABB) | Do whatever one can to reduce as much as possible consequences of a Botnet activity against the computer –victim of the attack |
| Monitoring, detection & studying of Bot (MDSBB) | Follow the situation in the field of security, detection of Bot and taking out of function and studying of Bots activities |
| Education of users (EOU) | Raise the level of security awareness with the users by permanent education |
| Legislative protection (LP) | Legislative measures-sanctions are to prevent the attackers from trying to carry the attacks out |

Description of measures and activities for each strategy, as well as their mutual comparison in accordance with several criteria is given further in the paper.

*A. Defense against infection by Bot*

Since Bot infection is carried out by means of viruses and worms, it is necessary to have total defense of computer or computer network, and this can be done by application of the following measures and activities:

- Permanently update operational system and other software; since Bots use security failures in operating system and other software, it is necessary to update them permanently, or install the patches and update [15]. Most operating systems and software have possibility of automatic updating, and it is necessary to turn this function on.

- Never click on attachments in e-mail before checking up who do sender is [16]. Since large number of Bots is spread by SPAMs in such a way that malicious code is sent in the attachment, it is necessary to check the sender before opening/running attachments. Even this procedure sometimes does not guarantee correctness of the message. The solution is to turn on the option for showing of file extension at the level of operating system. If you do not expect some message or you are not familiar with the source of the message, never run executive files with scripts (*.exe, *.com, *.bat, *.bin, *.vb, *.js, etc.).

- Turn off the support for scripting languages; Turning off the support for scripting languages such as JavaScript, VBScript and ActiveX or at least control of their usage, reduces the risk of Bot installing.

- Increase the level of security settings on the Internet Browser. Set the level of security settings on your Internet browser at the level before approval of JavaScript execution and this procedure will eliminate infection by Bot.

- Install Top-Rated protection software; it is necessary to install anti-virus, anti-Trojan and Personal Firewall in order to be protected against Bots. Of course, it is necessary to update the software permanently. For having easier updating and maintenance, some of integral software solutions such as Internet Security software (Kaspersky Internet Security or Norton Internet Security) can be installed.

- Limit the user's rights when you are Online; Limitation of the user's rights reduces possibility of Bot installing on the computer. This procedure can be carried out by creation of very low privileges level of user account which is used when using Internet.

- By defining of network Policy of the network administrators it is easy and simply to carry out the measures mentioned above, thus reducing the risk to which computers in the network are exposed to.

*B. Defense against Botnet attacks*

Web services/applications are in most cases attacked by Botnets. While creating the system, special attention should be paid to protection. Security must be built-in during each phase of the system development and this must be done on everyday basis. In order to be protected, it is necessary to obey the following rules:

- Usage of Intrusion of Prevention System (IPS); Intrusion Prevention Systems are devices which

monitor network activity in order to detect vilifications or undesirable activities in real time with the task to block or prevent them from acting [17], [19]. Figure 2 shows an example of Web service architecture with built-in IPS.

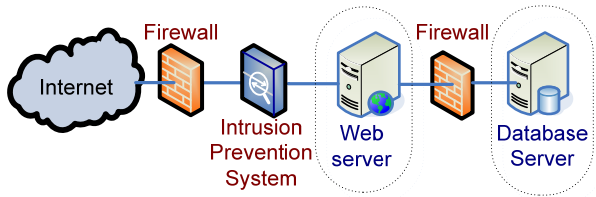

Figure 2. Example of a Web service arhitecture with IPS

IPSs function in the following way: when they detect attack or undesirable behavior, they reject undesirable packages and they let the other traffic go on. Although IPSs are similar to Firewall, their way of approaching the network and security system is fundamentally different. The contents can be filtered in several ways, and they are classified into different types in accordance with the way. The most important ones are: Content-based, Protocol Analysis and Rate-based. Content-based IPS controls the content of the package, looking for the sequence or signature needed for detection and prevention of already known kinds of attacks such as worm and Trojan infections. Protocol Analysis can decode application-layer network protocols such as HTTP or FTP. After the protocol is completely decoded, IPS analysis engine can detect strange behavior, i.e. unexpected values in the packages and reject the packages. Rate-based ones are primarily intended for prevention of Denial of Service (DoS) and Distributed Denial of Service (DDoS) attacks. They monitor and learn about normal behavior of the network. They follow the traffic in real time and compare it with the stored statistics. Based on the results, they can recognize abnormal rates of certain types of traffic such as TCP, UDP or ARP packages. The attacks can be identified when they exceed thresholds. The thresholds are dynamically adjusted depending on the part of day, day of the week etc, in accordance with the stored statistics.

- Usage of Transport Layer Security (TLS)/Secure Sockets Layer (SSL) protocol; Usage TLS/SSL protocol provides protected communication between the user and the service server. This reduces possibility for Botnet activity in the field Traffic Monitoring and this in the same time reduces compromising of the information.

- Correct coding of Web service/applications; Correct coding, without security failures, creates Web services/applications resistant to threats as well as to Botnets. If ready-made solutions are applied, it is necessary to undertake all the measures for permanent updating.

C. *Monitoring, detection and studying of Botnets*

All users of computers, but especially the system administrators, can detect presence and/or activities of Botnet and prevent it from influencing upon the computers by constant following of the activities on the computers. The monitoring activities must be carried out in order to achieve the objective mentioned here:

- Monitoring log files: This is one of the most important activities which must be performed every day. Studying of IDS log files, Firewalls, Mail Servers, DHCP Servers, IRC Servers, Proxy Servers etc. can notice intensifying of the traffic, and that can be a signal that Bot is in the network [20]. When it is noticed that the traffic is intensified, it is necessary to use the appropriate tools in order to identify subnets and/or computers which generate such traffic. Cooperation between Internet providers and system administrators provides conditions for finding and switching off the computers which are used as IRC Servers on the Internet. Special attention should be paid to Port 6667 (communication port of IRC protocol), as well as to the high number ports such as 31337 and 54321. In fact, all the traffic on the ports above 1024 should be blocked, except the ones which possibly use the applications within given organization. Since some Botnets use port 80 for communication with their Bots, special attention should be paid to the traffic by this port. Botnets usually update their zombies between 1:00 and 5:00 a.m., in the time when probably nobody monitors them. That's why the activities by port 80 should be checked every morning and, in accordance with the results, certain measures should be taken.

- To follow situation in the field of security, threats and countermeasures; Constant following of the situation in the field of security or, in other words, collecting of the latest information concerning security threats, their activities and countermeasures can provide the information necessary for upgrading of the system security.

- To look for and detect the threats: Using sophisticated techniques for looking for and detection of the threats can provide very important information and conclusions concerning preventive activities. One of the techniques is honeybots. Honeybots is a computer adjusted in such a way that it is an easy target for the attackers [18]. Their role is to become infected and enable the administrator to see exact source of the problems and to analyze the methods of the attack.

D. *Education of users*

Education of users (EOU) can be one of the strategies of defense against Botnets. Irrespectively of all measures and precautions undertaken for the system protection, the most efficient defense against Botnets is in the very user and his level of awareness. Constant education of users can raise the level of the users' responsibility and, in the same time, it raises the level of the computer security.

*E. Legislative protection*

Since the measures mentioned above are not carried out by all users of the Internet and having in mind the fact that different users have different attitudes toward the Internet, reality shows that defense against Botnets cannot be carried out only at IT level. Common acting of the Internet community and legislative-punishment policy in some countries have brought to significant reduction in number of some kinds of attacks. For example, these attacks almost disappeared in some countries where punishment policy is very strict (USA, EU countries, China, Japan, etc.) [21]. The punishments in these countries are either in form of big fines and/or several years in prison for the attackers. In the USA it is possible to spend even 10 years in prison for Denial of Service Attacks [22].

The problem lies in the fact that this issue is not treated in the same way in many countries and that's why appropriate legislative solutions haven't been defined yet. This caused "moving" of the servers from which the attacks are carried out to some other regions.

For the defense application in as many countries as possible, it is necessary to have common activities of the Internet communities and legislative institutions in order to define the ways of punishing and their realization. Besides, some legislative systems function on the basis of prohibition. It is necessary to follow the changes in the field of security, especially when new kinds of attacks are concerned, and to adjust the law regulations to them.

*F. Comparison of defense strategies*

The defense strategies mentioned so far can be mutually compared according to several criteria, as it is shown in the following tables (Table II, Table III, Table IV, Table V, Table VI, Table VII, Table VIII, Table XI and Table X).

TABLE II. WHO APPLIES THEM?

| Defense strategies | Response |
|---|---|
| DAIBB | All users of computers |
| DAABB | Developers, Web administrators |
| MDSBB | Administrators |
| EOU | All users of computers, individually, attending the courses |
| LP | State institutions in cooperation with the system administrators |

TABLE III. WHERE TO APPLY THE PROTECTION?

| Defense strategies | Response |
|---|---|
| DAIBB | On all computers |
| DAABB | With the service given on the Internet |
| MDSBB | At home, at work |
| EOU | At home, at work, at school |
| LP | The system institutions - courts |

TABLE IV. WHAT ARE THE ADVANTAGES?

| Defense strategies | Response |
|---|---|
| DAIBB | Protect computer against Bot infection |
| DAABB | Protect the system against Botnet attacks |
| MDSBB | Getting some new information about Bot characteristics |
| EOU | Raising of the user's awareness concerning security |
| LP | Prevention of potential attackers before they launch the attack |

TABLE V. WHAT ARE DISADVANTAGES?

| Defense strategies | Response |
|---|---|
| DAIBB | Mostly depends on the user's responsibilty |
| DAABB | Time is necessary for "studying" |
| MDSBB | It acts only after Bots activities have already been noticed |
| EOU | Some users are not willing to refresh and spread their knowledge concerning security |
| LP | Different ways of punishing |

TABLE VI. DEPENDENCE ON SOME OTHER STRATEGY

| Defense strategies | Response |
|---|---|
| DAIBB | Education of users |
| DAABB | Protection against infection by Bot, Education of users |
| MDSBB | Education of users (administrators) |
| EOU | It doesn't depend on any other strategy |
| LP | It doesn't depend on any other strategy |

TABLE VII. PRICE

| Defense strategies | Response |
|---|---|
| DAIBB | Low |
| DAABB | Price of the prevention system and its maintenance- high |
| MDSBB | Price of the administrators' hours |
| EOU | Relatively low |
| LP | Relatively low |

TABLE VIII. DIFFICULTY OF IMPLEMENTATION

| Defense strategies | Response |
|---|---|
| DAIBB | Very simple |
| DAABB | Relatively simple |
| MDSBB | Relatively simple |
| EOU | Relatively difficult |
| LP | Relatively simple |

TABLE IX. LEVEL OF AUTOMATIZATION

| Defense strategies | Response |
|---|---|
| DAIBB | High level |
| DAABB | High level |
| MDSBB | Relatively high level |
| EOU | Doesn't exist |
| LP | Doesn't exist |

TABLE X. THE TIME NEEDED FOR GETTING THE RESULTS

| Defense strategies | Response |
|---|---|
| DAIBB | Immediately after application |
| DAABB | Time needed for "studying" |
| MDSBB | Depends on the tools and qualification of personnel |
| EOU | Immediately after the knowledge acquirement |

| Defense strategies | Response |
|---|---|
| LP | Time needed for creation and acquirement of new and/or changing of the exisitng solutions |

It is obvious that these strategies are complementary ones and that application of all the strategies is the best solution for defense against Bot Botnets. However, application of all the strategies is the most expensive solution and it is not realistic to expect that all the Internet users will apply them. On the other hand, application of Defense strategy against Bot is a minimally acceptable solution.

## VII. CONCLUSION

Botnets are nowadays very important challenge for the Internet community. Their activities cause a lot of problems for the Internet users, system administrators, providers, etc. The threats put upon the computers and systems by Botnets require efficient defense strategies.

By defining the strategies for defense against Botnets, this paper offers a new attitude towards the possible solutions of this problem. The paper gives description of the measures and gives proposal of activities for carrying out of each strategy, as well as comparison of the strategies according to several criteria. Applying strategies contributes to significant results in defense against Botnets.

AUTHORS PROFILE

Srdjan Stanković is a MSc student at the Faculty of Organizational Sciences, University of Belgrade. He is working at Ministry of Defense as a Software Analyst. His interests are Web application security, security of computer systems, applied information technologies.

Dejan Simić, PhD, is a professor at the Faculty of Organizational Sciences, University of Belgrade. He received the B.S. in electrical engineering and the M.S. and the Ph.D. degrees in Computer Science from the University of Belgrade. His main research interests include: security of computer systems, organization and architecture of computer systems and applied information technologies.